\documentclass[11pt,twoside,]{article}
\usepackage{asp2006}
\usepackage{epsf}
\usepackage{graphicx}

\begin{document}

\title {Simulations of Buoyant Plumes In Solar Prominences}
\author {N.  Hurlburt and T. Berger}
\affil{Lockheed Martin Advanced Technology Center, ADBS/B252, 3251 Hanover Street, Palo Alto, CA 94304 USA}

\date{Draft: 11 September 2008}

\begin{abstract}
Observations of solar prominences reveal a complex, dynamic flow field within them. The flow field within quiescent prominences is characterized by long ``threads'' and dark ``bubbles'' that fall and rise (respectively) in a thin sheet. The flow field in active prominences display more helical motions that travel along the axis of the prominence. We explore the possible dynamics of both of these with the aid of 2.5D MHD simulations. Our model, compressible plasma possesses density and temperature gradients and resides in magnetic field configurations that mimc those of a solar prominence. We present results of various configurations and discuss the nonlinear behavior of the resulting dynamics.
\end{abstract}

\section{Introduction}
Recent observations of solar prominences have revealed a complex, dynamic flow field within them. The flow field within quiescent prominences is characterized by long ``threads'' and dark ``bubbles'' that fall and rise (respectively) in a thin sheet. The flow field in active prominences display more helical motions that travel along the axis of the prominence. We investigate these behaviors with the aid of axisymmetric MHD simulations.

Flows within prominences may be influenced by double diffusive instabilities, which have been extensively studied for the role they may play in the solar dynamo \cite{HughesWeiss95}. In this case the magnetic fields provide buoyancy to a rising a fluid parcel through the magnetic pressure even if the thermal background is stably stratified. This can occue when  ${d{B^2/\rho}\over dz} < 0$, where $B$ is the horizontal magnetic field strength, $\rho$ is the density and $z$ is the height. In other words, when  the Alf\`ven speed in a fluid layer decreases with height, magnetic "fingering" instabilities are possible. 

It is quite likely that such conditions can be met in prominences. If prominences are plasma trapped about the solar photosphere by a helical field whose axis is aligned to the horizontal. then the region below the axis of the flux tube will indeed be a location where the field decreases with height. If this corresponds to a region with relatively large density scale heights, fingering instabilities are possible. Here we concoct a simple axisymmetric model of a prominence to investigate the feasibility and possible signatures of any such dynamics. We present our model in the next section, display our numerical results in the next and close with a discussion and conclusion.

\section{Model}
\label{sec:model}

Partial differential equations describing compressible MHD 
are solved in an axisymmetric cylindrical geometry, using a numerical code 
developed for this purpose  \citep{HurlburtRucklidge00}. \citet{BothaEA08} extend this model to include azimuthal flows and fields. Here we modify the initial state and background stratification to mimic a horizontal, helical flux tube and consider the dynamics of buoyancy instabilities.

The computational domain is an axisymmetric cylinder of inner, outer radius $\Gamma_1, \Gamma_2$, 
situated in the $(r,z)$ plane. At the radial boundaries boundary the magnetic field is imposed to be consistent with its initial condition; the temperature gradient is chosen to be zero and the horizontal boundaries are impenetrable and stress free. The inner and outer walls are perfect conductors and field lines are tied on the surfaces $z=0,1$. The plasma is triggered by starting the quiet, hydrostatic equilibrium perturbed by a hot blob in pressure equilibrium near the outer boundary. The initial temperature and density profiles in the radial ($r$) 
direction are $T=T_0,$ $\rho=\rho_0 e^{g r^2/2},$ $B_\phi=B_0 (r/ \Gamma_2)^2 \sin{\theta},$ and $B_z=B_0 \cos{\theta}. $ Here, the $0$ subscript defining the quantity at the outer edge of the cylinder ($r=\Gamma_2$);
$\theta$ represents the initial angle of the field twist  along the outer wall;
and the gravitational force constant $g^{-1}=H T_0 \Gamma_2$ where $H$ is the pressure scale height at $r=\Gamma_2$. 

The physical quantities are dimensionless, with the length scaled by the length of cylinder, velocity scaled by the sound speed in it and temperature, magnetic field, density, and pressure all scaled to their initial values at $r=0$.
\section{Numerical results}
\label{sec:numres}
\begin{figure}
  \centering
   \includegraphics[width=3.5truein]{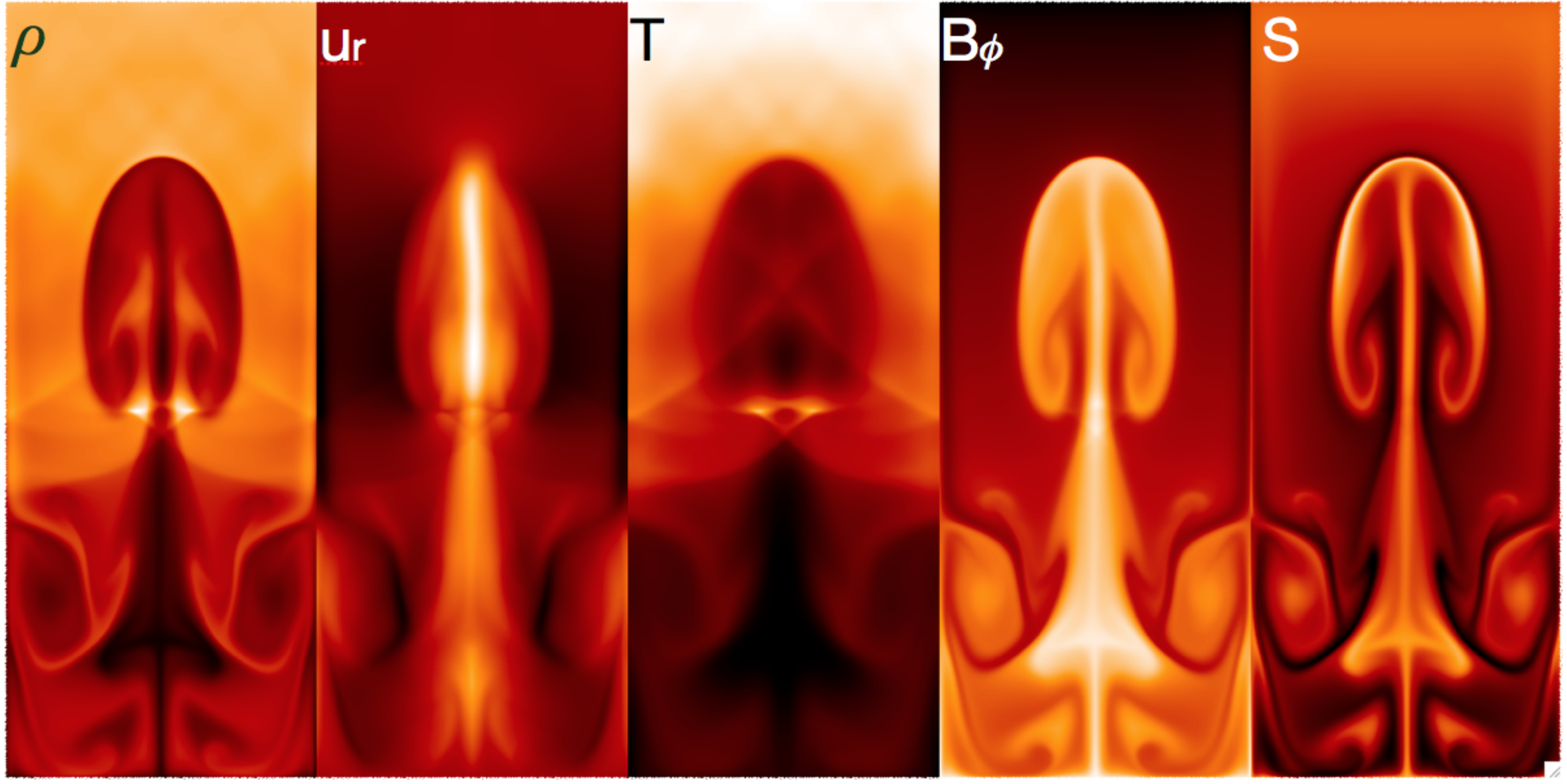} 
   \caption{The detailed properties of one buoyant plume at mid-rise shows that the buoyancy due to the effect of magnetic pressure rather than temperature. The result is a cool, low-density object that would appear as a dark feature Hinode images. The radial (vertical) velocity $u_r$ and entropy $S $ are also shown.}
   \label{fig:plume}
\end{figure}
\begin{figure}
   \centering
 \includegraphics[width=3.0truein]{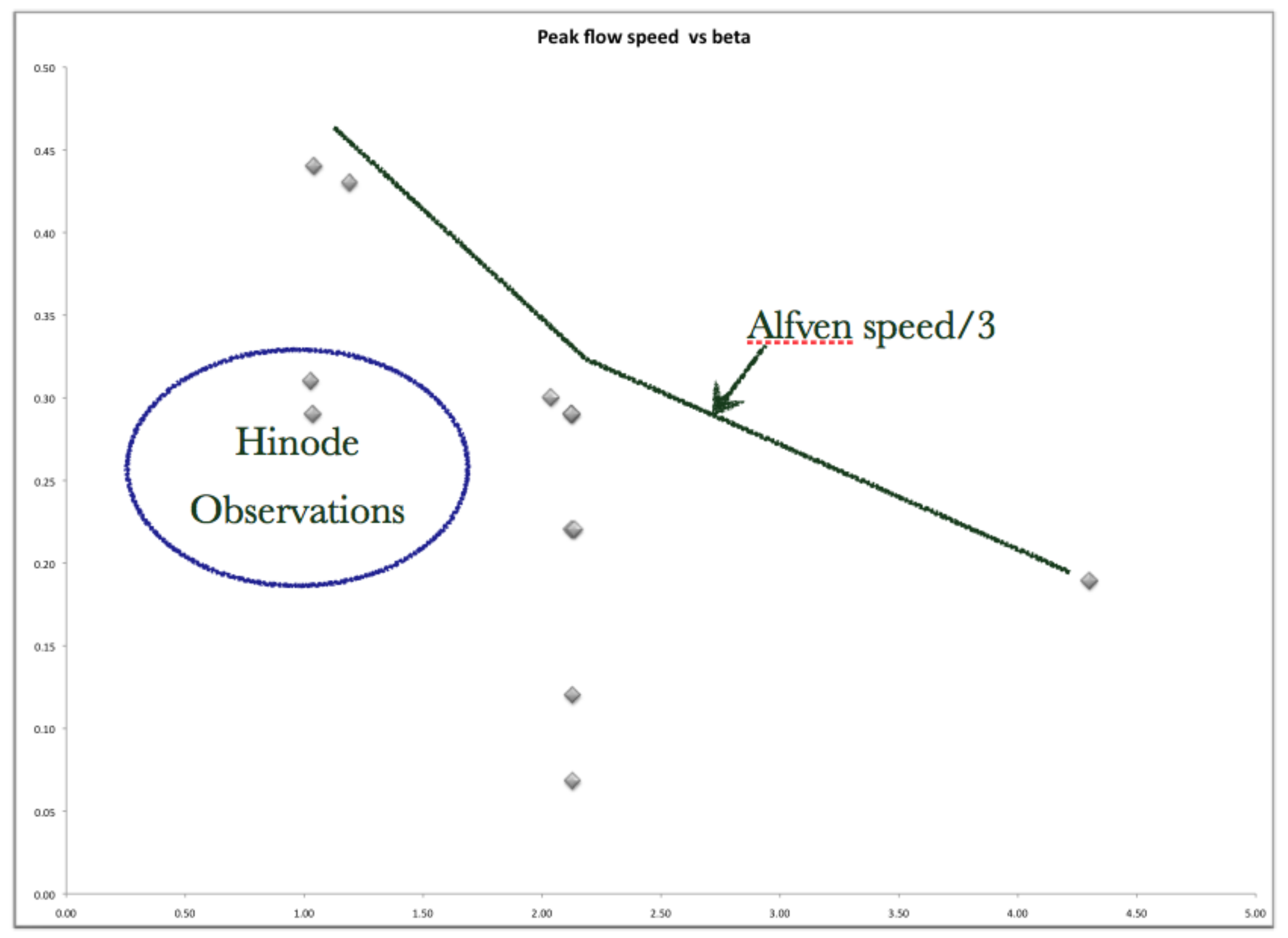}  
   \caption{A series of simulations with varying plasma-$\beta$ reveals a peak velocity that scales with Alf\`ven speed. Hinode observations (scaled to these non-dimensional units) are consistent with these speeds.}
   \label{fig:survey}
\end{figure}
With a purely azimuthal field, the plumes form and rise with classic plume profile as seen in Figure \ref{fig:plume}. The buoyancy driving from the initial thermal pulse  is replaced by the magnetic pressure and produces a strong upflow as seen in $u_r$. The  maximum speed of this flow as a function of initial plasma-$\beta$ is displayed in Figure \ref{fig:survey}. The flows can reach speeds of up to 1/3 of the ambient Alf\`ven speed, for all values of $\beta$. These values are consistent with those observed by Hinode/SOT \citep{Berger/Shine/etal:2008}.

\begin{figure}
   \centering
   \includegraphics[width=3.0truein]{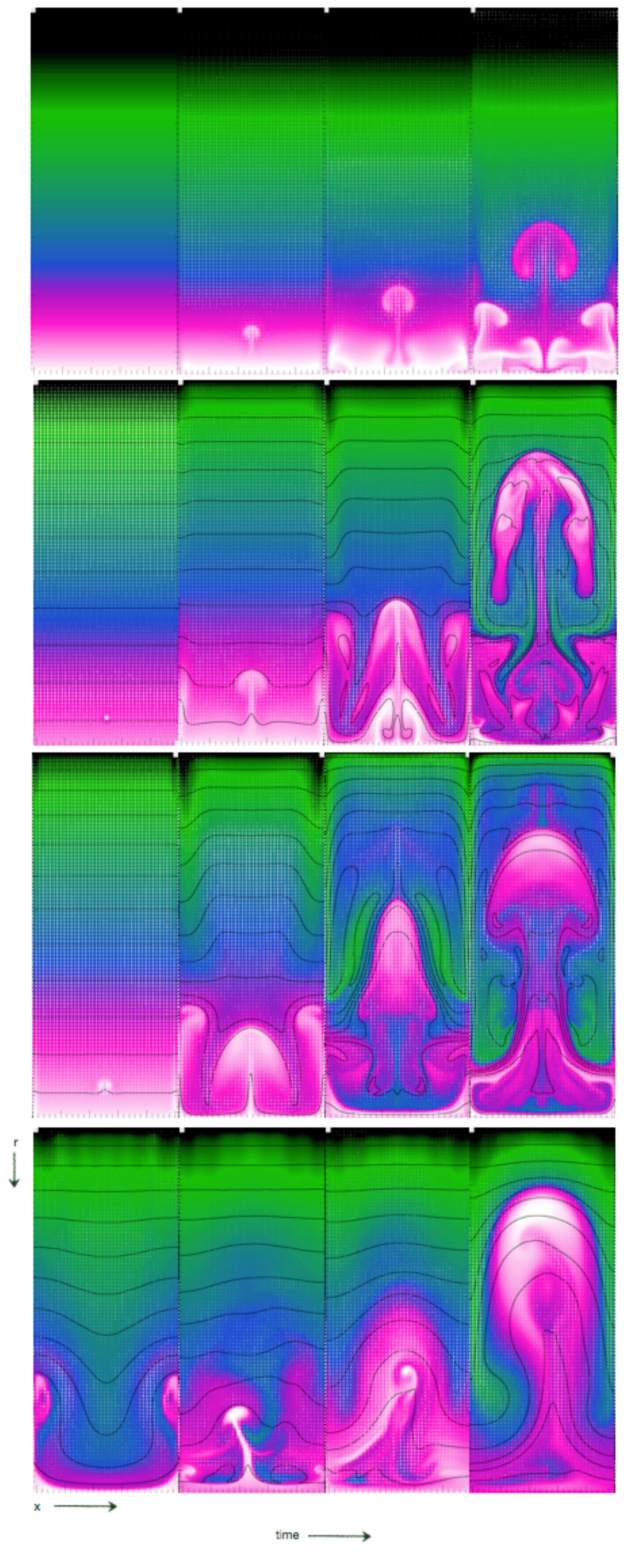} 
   \caption{A series of simulations in time with plasma-$\beta$ near unity and increasing field twist. The background color is the azimuthal field strength, the lines are the axial field and the small white arrows show the flow field in the plane. The top panel is a purely azimuthal field ($\phi=90$). The twist increases as $\phi=89.4, 88.8$ and  $87.0$ degrees respectively.}
   \label{fig:plumes}
\end{figure}

When an axial component of the field is introduced (decreasing the twist) the dynamics becomes richer. A series of simulations with decreasing twist is displayed in Figure \ref{fig:plumes}. The solutions display axial Alf\`ven waves and the instability transitions from a direct, growing solution to an overstable oscillation that eventually bifurcates to a direct mode. At smaller twists the instabilities are not properly captured within this 2.5D model.

\section{Summary}
We have presented a series of simulations of magnetic flux bubbles rising into an idealized prominence. The bubbles, driven by magnetic buoyancy, display morphological characteristics similar to those observed by Hinode/SOT \citep{Berger:2010} and can attain rise speeds that are consistent with these observations. The maximum speed increases in proportion to the imposed Alf\`ven speed, even into the regime where the plasma-$\beta$ is below unity.

This work has been supported by NASA through contract NNM07AA01C.


\label{lastpage}

\end{document}